\documentclass[%
 reprint,
 prl,
superscriptaddress,
 amsmath,amssymb,
 aps,
]{revtex4-2}

\usepackage{graphicx}
\usepackage{dcolumn}
\usepackage{bm}
\usepackage{hyperref}

\usepackage{color}

\begin{document}

\preprint{APS/123-QED}

\title{Information-Geometric Signatures of Nonconservative Driving}

\author{Andrea Auconi}
\email{andrea.auconi@gmail.com}
\affiliation{%
 Ca’ Foscari University of Venice, DSMN - via Torino 155, 30172 Mestre (Venice), Italy
}%

\author{Sosuke Ito}
\thanks{sosuke.ito@ubi.s.u-tokyo.ac.jp}
\affiliation{Department of Physics, The University of Tokyo, 7-3-1 Hongo, Bunkyo-ku, Tokyo 113-0033, Japan}
\affiliation{Universal Biology Institute, The University of Tokyo, 7-3-1 Hongo, Bunkyo-ku, Tokyo 113-0033, Japan}

\date{\today}

\begin{abstract}
We propose an information-geometric signature of nonconservative driving that detects violations of detailed balance using the Kullback--Leibler divergence and the Fisher information. For Markov jump processes satisfying detailed balance, we show that, near equilibrium, the acceleration of the Kullback--Leibler divergence relative to the equilibrium state is given by twice the Fisher information with respect to time. In contrast, for relaxation toward a nonequilibrium steady state, this relation is generally violated even near the steady state. We refer to the resulting discrepancy as the relaxation gap and derive a lower bound on the steady-state entropy production rate in terms of this gap. We demonstrate that this bound is particularly tight for networks with simple cyclic topologies. Finally, we show that analogous relations and bounds hold for Fokker--Planck dynamics.
\end{abstract}

                              
\maketitle

Determining whether stochastic dynamics satisfies detailed balance is fundamental for characterizing nonequilibrium behavior. When detailed balance is broken, nonconservative driving can sustain probability currents even in the nonequilibrium steady state (NESS), leading to a positive entropy production rate that quantifies dissipation~\cite{schnakenberg1976network,seifert2025stochastic}. Systems without detailed balance have been studied in terms of oscillatory behavior~\cite{schnakenberg1976network,barato2017coherence,ohga2023thermodynamic,uhl2019affinity,xu2025thermodynamic}, relaxation speed~\cite{diaconis2000analysis,bremaud2013markov,turitsyn2011irreversible,suwa2010markov,ichiki2013violation,kaiser2017acceleration,shiraishi2019information,kolchinsky2024thermodynamic,kolchinsky2026generalized}, response~\cite{prost2009generalized,mandal2016analysis,owen2020universal,aslyamov2025nonequilibrium}, and stability~\cite{glansdorff1964general,
schlogl1971stability,glansdorff1974thermodynamic,de1975glansdorff,schnakenberg1976network,maes2015revisiting,ito2022information,tome2025irreversible,auconi2025nonequilibrium}. In particular, in Markov-chain Monte Carlo methods, dynamics without detailed balance are known to accelerate convergence to a target distribution~\cite{bremaud2013markov,turitsyn2011irreversible,suwa2010markov,ichiki2013violation}. Thermodynamic trade-off relations~\cite{kolchinsky2024thermodynamic} inspired by thermodynamic uncertainty relations~\cite{barato2015thermodynamic,horowitz2020thermodynamic} have also motivated studies on the connection between relaxation speed and steady-state dissipation.

Previous studies have often investigated the effects of detailed balance violation by fixing the steady-state distribution and comparing conservative dynamics satisfying detailed balance with nonconservative dynamics violating it~\cite{bremaud2013markov}. While these approaches clarify how nonequilibrium currents modify relaxation, they do not directly provide a criterion for detecting detailed balance violation from the relaxation behavior of a single system. Here, we develop such a criterion from an information-geometric perspective~\cite{amari2016information, crooks2007measuring,sivak2012thermodynamic,ito2024geometric}. Information geometry characterizes dynamical changes in probability distributions through quantities such as the intrinsic speed~\cite{ito2018stochastic,ito2020stochastic,nicholson2020time}, defined as the square root of the Fisher information with respect to time, and the Kullback–Leibler divergence~\cite{schlogl1971stability,amari2016information} relative to the stationary distribution. Although these quantities have been used to analyze the stability of NESSs~\cite{ito2022information} and formulate thermodynamic trade-off relations near NESSs~\cite{auconi2025nonequilibrium}, how they are constrained by detailed balance during relaxation remains unclear.

We show that, for Markov jump processes satisfying detailed balance, the acceleration of the Kullback–Leibler divergence relative to the equilibrium state is given by twice the squared intrinsic speed near equilibrium. In contrast, for relaxation toward the NESS, this relation is generally violated, and we refer to the resulting discrepancy as the relaxation gap. We further derive, near the NESS, a bound on the relaxation gap in terms of the NESS entropy production rate, linking this information-geometric signature of nonconservative driving to the dissipation required to maintain the NESS. Our results apply to both discrete-state Markov jump processes and continuous-state Fokker–Planck dynamics. We also show that the bound can be fully saturated in networks with simple cyclic topologies, confirming its tightness.

\paragraph*{Setup.}

Consider a system with $n$ allowed microstates or nodes, and denote by $\boldsymbol{p}\equiv \left[ p_1(t), ... , p_n(t) \right]$ the probability vector in this discrete space.
We consider its time evolution as a Markov jump process (MJP) \cite{van1983stochastic}, 
\begin{equation}\label{MJP}
    d_t p_i = \sum_{j} M_{ij} p_j,
\end{equation}
which we assume to be ergodic \cite{van1983stochastic},
where $M_{ij}\geq 0$ is the constant transition rate from node $j$ to node $i (\neq j)$, $ -M_{ii}\equiv \sum_{j (\neq i)} M_{ji} > 0$ is the escape rate of node $i$, and all sums over $j$ run from $1$ to $n$. In matrix notation, this is written as $d_t \boldsymbol{p} = M \boldsymbol{p}$.
We assume transitions to be locally reversible, meaning $M_{ji}>0$ if $M_{ij}>0$. 
The anti-symmetric net current on the edge $j\rightarrow i$ is defined as $J_{ij} \equiv M_{ij} p_j - M_{ji}  p_i$, 
and the dynamical activity~\cite{maes2020frenesy} is defined as $A_{ij} \equiv M_{ij} p_j + M_{ji}  p_i$, 
which leads to the expression $d_t p_i= (1/2)\sum_{j} (A_{ij} + J_{ij})$.
Let us also define the thermodynamic force,  $F_{ij} \equiv \ln [(M_{ij} p_j)/(M_{ji}  p_i)  ]$,
and note the relation $A_{ij}= J_{ij} \coth (F_{ij} / 2)$ valid for $A_{ij} > 0$ ($i \neq j$).

Denote by $\boldsymbol{p}^*$ the NESS distribution, which is defined by
$d_t \boldsymbol{p}\vert_{\boldsymbol{p}=\boldsymbol{p}^*} = \boldsymbol{0}$, and whose uniqueness is guaranteed by the ergodicity of the dynamics through the Perron-Frobenius theorem \cite{van1983stochastic}. Explicitly, from Eq.~\eqref{MJP}, the steady state condition is written as $\sum_{j} M_{ij} p^*_j=0$.

The NESS entropy production rate \cite{schnakenberg1976network,seifert2025stochastic} is defined as
\begin{equation}\label{EP}
    \sigma^* \equiv \frac{1}{2}\sum_{i,j} J^*_{ij} F^*_{ij},
\end{equation}
where the superscript $^*$ denotes quantities evaluated at the steady state (e.g., $J_{ij}^* = J_{ij}\vert_{\boldsymbol{p}=\boldsymbol{p}^*}$ and $F_{ij}^* = F_{ij}\vert_{\boldsymbol{p}=\boldsymbol{p}^*}$) and therefore constant in time.
The detailed balance condition, $\forall (i,j)$, $J^*_{ij}=0$, corresponds to the equilibrium case $\sigma^* = 0$. 
Let us further define averages with respect to the steady-state vector as $\langle f \rangle\equiv\sum_i p^*_i f_i$ for any test vector $\boldsymbol{f} =[f_1, \dots, f_n]$.


\paragraph*{Small perturbations.}

Let us define the perturbation vector $\boldsymbol{\phi}=[\phi_1, \dots, \phi_n]$ around the steady state as
\begin{equation}
    p_i = p^*_i e^{\phi_i},
\end{equation}
with normalization $\langle e^\phi \rangle = 1$.
Assume this perturbation to be small, meaning $\max_i |\phi_i| \ll 1$, so that its dynamics [Eq.~\eqref{MJP}] can be linearized to
\begin{equation}\label{d_t phi}
    d_t \phi_i = \frac{1}{2 p^*_i} \sum_{j} \left(A^*_{ij} +J^*_{ij}\right) \phi_j ,
\end{equation}
where we used $\sum_j M_{ij} p^*_j =0$, see the Supplementary Materials (SM) for details of the calculations. This expression explicitly decomposes the linearized dynamics into equilibrium and nonequilibrium parts.

    The Kullback-Leibler (KL) divergence \cite{amari2016information} of a distribution $\boldsymbol{p}$ relative to another distribution $\boldsymbol{q}$ is defined as $D[\boldsymbol{p}||\boldsymbol{q}] \equiv \sum_i p_i \ln \left( {p_i}/{q_i} \right)$. This quantity $D [\boldsymbol{p} || \boldsymbol{p}^*]$ is zero only when $\boldsymbol{p}= \boldsymbol{p}^*$.
 In the weak-perturbation limit $\max_i |\phi_i| \rightarrow 0$, considered hereafter and denoted by $\boldsymbol{p}\simeq \boldsymbol{p}^*$, we obtain
\begin{equation}\label{def D}
    D [\boldsymbol{p} || \boldsymbol{p}^*] = - \langle \phi \rangle= \frac{1}{2} \langle \phi^2 \rangle  \geq 0,
\end{equation}
where the normalization implies $\langle \phi \rangle = -\langle \phi^2 \rangle/2 +\mathcal{O}(\langle \phi^3 \rangle)$.

Let us study the stability of the NESS from the time evolution of the divergence,
\begin{equation}\label{stability}
    d_t D[\boldsymbol{p} || \boldsymbol{p}^*]   = \frac{1}{2}\sum_{i,j} A^*_{ij} \phi_i \phi_j \leq 0,
\end{equation}
where we used the symmetry $J^*_{ij}=-J^*_{ji}$, and the inequality follows from the steady-state dynamical activity $A^*$ being the negative of a weighted graph Laplacian matrix which is known to be negative-semidefinite \cite{chung1997spectral}. Indeed, it is symmetric $A^*_{ij}=A^*_{ji}$, and the steady-state condition implies that the sum over any row or column is zero, $\sum_{i} A^*_{ij} = \sum_{j} A^*_{ij} = 0$.
Eq.~\eqref{stability} means that, by evaluating the dynamics at second order around the steady state [Eq.~\eqref{def D}], the Glansdorff-Prigogine criterion for stability \cite{glansdorff1974thermodynamic,schnakenberg1976network, maes2015revisiting, ito2022information} is satisfied.
This establishes $D[\boldsymbol{p}||\boldsymbol{p}^*]$ as a valid Lyapunov function whose monotonic decrease governs the approach to the steady state~\cite{schlogl1971stability,de1975glansdorff}. Furthermore, the nonnegative quantity $-d_t D[\boldsymbol{p} || \boldsymbol{p}^*] (\geq 0)$ is known as the Hatano-Sasa excess (or nonadiabatic) entropy production rate~\cite{hatano2001steady,esposito2010three,dechant2022geometric}.


\paragraph*{Intrinsic speed and relaxation.}
We first consider the intrinsic speed of the dynamics $v_{\rm info}(\boldsymbol{p})$.  The Fisher information metric \cite{amari2016information} with respect to time defines the squared intrinsic speed~\cite{ito2018stochastic, ito2020stochastic},
\begin{equation}\label{ds2_def}
    \left( v_{\rm info} (\boldsymbol{p})\right)^2 \equiv \sum_i \frac{(d_t p_i)^2}{p_i} = \sum_i p_i (d_t \ln p_i)^2 \geq 0. 
\end{equation}
This intrinsic speed is also given by $( v_{\rm info} (\boldsymbol{p}(t)))^2= d_{\tau}^2 D[\boldsymbol{p}(t+\tau)||\boldsymbol{p}(t) ] \vert_{\tau =0}$.
The expression near the steady state $\boldsymbol{p}\simeq \boldsymbol{p}^*$ is given by
\begin{equation}\label{ds2_def}
    \left( v_{\rm info} (\boldsymbol{p}) \right)^2 =  \langle (d_t\phi)^2 \rangle .
\end{equation}

\begin{figure}[tbp]
    \centering
    \includegraphics[width=0.98\columnwidth]{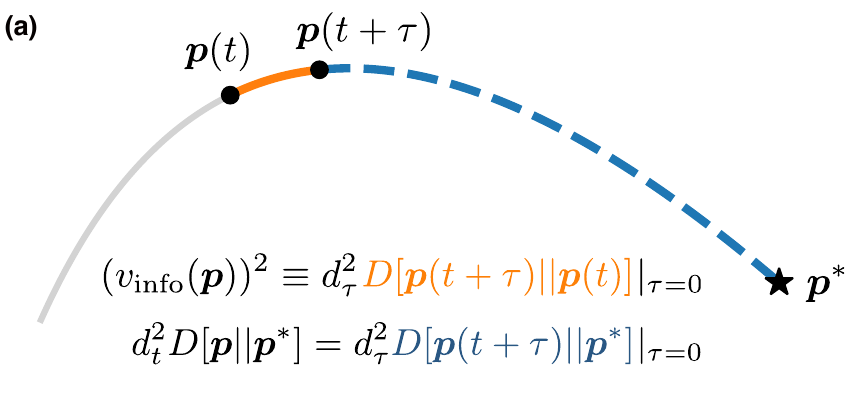}
    
    \vspace{0.4cm} 
    
    \includegraphics[width=\columnwidth]{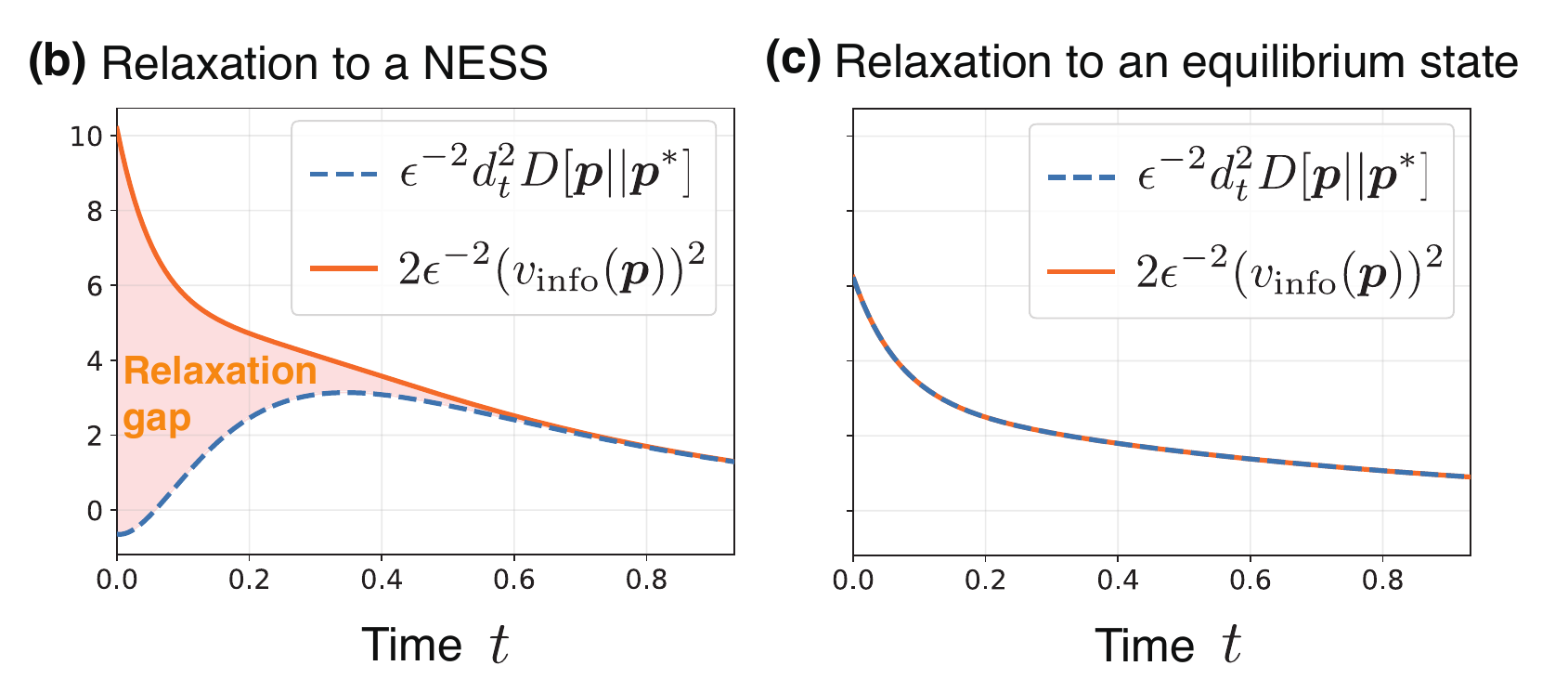}
    
    \caption{(a) Schematic representation of a probability trajectory $\boldsymbol{p}(t)$ towards the steady state $\boldsymbol{p}^*$.  By considering a small time interval $\tau\rightarrow 0$, the local intrinsic speed $v_{\rm info} (\boldsymbol{p})$ and the relaxation acceleration $d^2_t D [\boldsymbol{p} || \boldsymbol{p}^*]$ both originate from the curvature $d^2_\tau$ of the KL divergence of $\boldsymbol{p}(t+\tau)$ relative to $\boldsymbol{p}(t)$ and $\boldsymbol{p}^*$, respectively.
    (b) An example of relaxation to a nonequilibrium steady state (NESS).  There is a relaxation gap, i.e., $2 (v_{\rm info} (\boldsymbol{p}))^2\neq d^2_t D [\boldsymbol{p} || \boldsymbol{p}^*]$. Here, $\epsilon = 10^{-4}$ ($\propto \phi_i$) is the amount of perturbation from the steady state.
    (c) An example of relaxation to an equilibrium state. We use the corresponding equilibrium rate $\widetilde{M}_{ij} =A^*_{ij}/(2 p^*_j)$ (see the SM), where currents are eliminated while preserving the steady-state distribution. There is no relaxation gap, i.e., $2 (v_{\rm info} (\boldsymbol{p}))^2= d^2_t D [\boldsymbol{p} || \boldsymbol{p}^*]$.}
    \label{fig:fig1}
\end{figure}

We next consider the relaxation toward the steady state.
During the relaxation process to the steady state, the relaxation acceleration is represented by
$d_t^2 D[\boldsymbol{p}||\boldsymbol{p}^*]$. Near the steady state $\boldsymbol{p}\simeq \boldsymbol{p}^*$, this is given by 
\begin{equation}\label{second derivative}
    d^2_t D [\boldsymbol{p} || \boldsymbol{p}^*]  
    =\sum_{i,j}  A^*_{ij} \phi_i d_t\phi_j,
\end{equation}
which depends on the steady-state currents $J^*$ through $d_t \boldsymbol{\phi}$. In Ref.~\cite{auconi2025nonequilibrium}, this observable was used to bound the steady-state entropy production by comparison with the corresponding equilibrium dynamics, see the SM for an extension to MJPs.

Figure~\ref{fig:fig1}, together with the SM, illustrates how  $(v_{\rm info} (\boldsymbol{p}))^2$, like the relaxation acceleration $d^2_t D[\boldsymbol{p} || \boldsymbol{p}^*]$, also derives from the curvature of a KL divergence.



Remarkably, there is a relation between intrinsic speed and relaxation acceleration, which constitutes the first main result of this Letter. Near the steady state $\boldsymbol{p}\simeq \boldsymbol{p}^*$, from Eqs.~\eqref{d_t phi}, \eqref{ds2_def} and \eqref{second derivative} we obtain the following identity
\begin{equation}\label{symmetry breaking}
    d^2_t D [\boldsymbol{p} || \boldsymbol{p}^*] - 2 \left( v_{\rm info} (\boldsymbol{p}) \right)^2 = \sum_{i,j}  J^*_{ij}  \phi_i d_t\phi_j,
\end{equation}
which imposes constraints on the behavior of relaxation toward the equilibrium state in systems that satisfy the detailed balance condition. In fact, for equilibrium systems, where $\forall (i,j)$, $J^*_{ij}=0$, Eq. \eqref{symmetry breaking} implies the identity $
d_t^2 D[\boldsymbol{p} || \boldsymbol{p}^*]=2\left( v_{\rm info} (\boldsymbol{p}) \right)^2$, meaning that the acceleration is fully determined by the intrinsic speed. 
We refer to the left-hand side of Eq. \eqref{symmetry breaking}, $d^2_t D [\boldsymbol{p} || \boldsymbol{p}^*] - 2 \left( v_{\rm info} (\boldsymbol{p}) \right)^2$, as the relaxation gap. Therefore, if this relaxation gap is nonzero, it indicates that the system does not satisfy the detailed  balance condition. Thus, a nonzero relaxation gap for a given perturbation is a sufficient signature of nonconservative driving. The relaxation gap vanishes when $d_t \phi_j=0$, as occurs as the system approaches the NESS.

\paragraph*{Relaxation gap and entropy production.}
Based on the relaxation gap identity [Eq.~\eqref{symmetry breaking}], we derive a new thermodynamic bound that links the behavior of relaxation to the NESS and the entropy production rate at the NESS. This is our second main result, and its derivation is given at the end of this Letter,
\begin{equation}\label{MJP bound}
         \sigma^*\geq \mathcal{B} \equiv \frac{ \kappa^{-1} \left( d^2_t D[\boldsymbol{p}||\boldsymbol{p}^*] -  2\left( v_{\rm info} (\boldsymbol{p}) \right)^2 \right)^2}
         { 2 D[\boldsymbol{p}||\boldsymbol{p}^*] (v_{\rm info} (\boldsymbol{p}))^2 - \left( d_t D[\boldsymbol{p}||\boldsymbol{p}^*] \right)^2 },
\end{equation}
where $\kappa$ is the fastest mixing rate~\cite{kolchinsky2024thermodynamic} 
\begin{equation}\label{kappa}
    \kappa \equiv \max_{(i,j)}  \frac{A^*_{ij}}{2 p^*_i p^*_j} > 0 ,
\end{equation}
which characterizes the maximum rate at which probability can flow between different regions of the state space~\cite{kolchinsky2024thermodynamic}. Note that $\mathcal{B}\geq 0$, meaning that Eq. \eqref{MJP bound} is a refinement of the second law of thermodynamics.

Because the bound must hold for arbitrary perturbations, $\kappa$ represents a worst-case choice in which relaxation proceeds through the fastest channel. This limits the tightness of the thermodynamic bound, but it enables us to encapsulate the microscopic details into this single steady-state quantity $\kappa$.



\begin{figure}
\centering
    \includegraphics[scale=0.6]{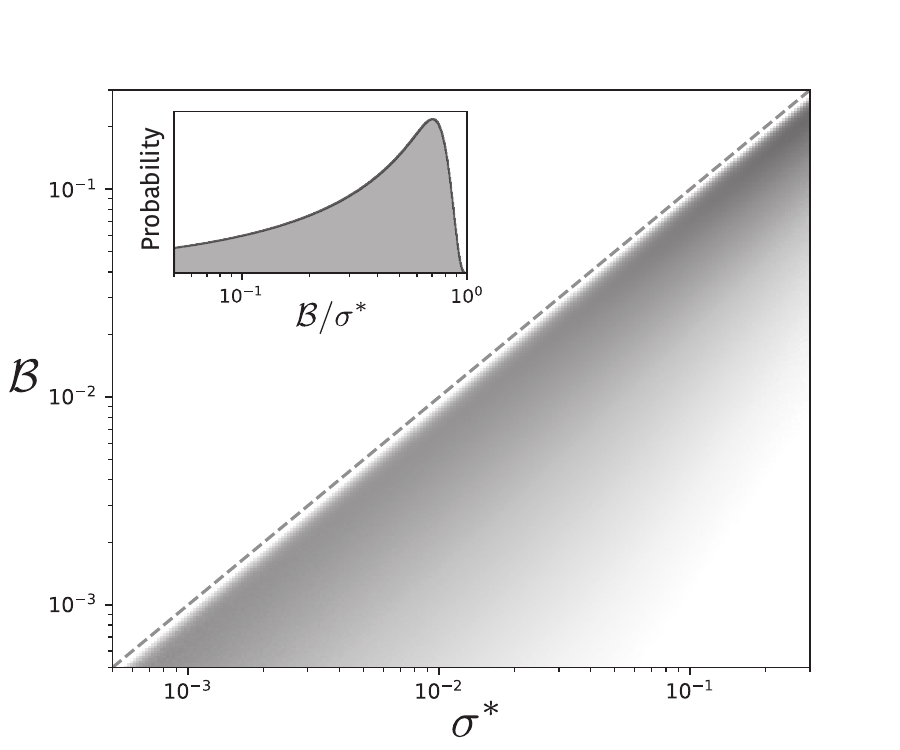}
        
\caption{Numerical analysis of the tightness for the bound \eqref{MJP bound}.
We display the conditional distributions of the bound given the entropy production rate, sampled from an ensemble of $10^8$ random rate matrices and perturbation vectors with $n=4$ states. The diagonal dashed line indicates the saturation limit.
The inset depicts the distribution of the tightness ratio $\mathcal{B}/\sigma^*$. Transition rates $M_{ij}$ were drawn independently from a log-normal distribution, $\ln M_{ij} \sim \mathcal{N}(m, 0.2)$, with $m$ chosen such that the mean rate is $\mathbb{E} [M_{ij} ]=1$ for $i\neq j$.}
    \label{fig:Tightness}
\end{figure}

We have analyzed the tightness of the inequality \eqref{MJP bound}, first analytically for a cycle model that reaches full saturation, and then numerically on random realizations of the transition matrix and perturbation, see Fig. \ref{fig:Tightness}.

\paragraph*{Uniform biased cycle.}

To gain intuition, we analytically study a simple cyclic topology. 
Consider a system of $n$ discrete states arranged on a ring structure with periodic boundary conditions with uniform transition rates $k_+$ and $k_-$ in the forward and backward directions, respectively,
\begin{equation}
M_{ij} = 
\begin{cases} 
k_+ & \text{if } i = j+1 \pmod n \\
k_- & \text{if } i = j-1 \pmod n \\
-(k_+ + k_-) & \text{if } i = j \\
0 & \text{otherwise}
\end{cases}
\end{equation}
The steady-state probability distribution is uniform due to symmetry, $p^*_i = 1/n $, and the linearized time evolution is written as $d_t\phi_i = k_+ \phi_{i-1} + k_- \phi_{i+1} - (k_+ +k_-) \phi_i$.

\begin{figure}[htbp]
    \centering
\includegraphics[width=0.48\textwidth]{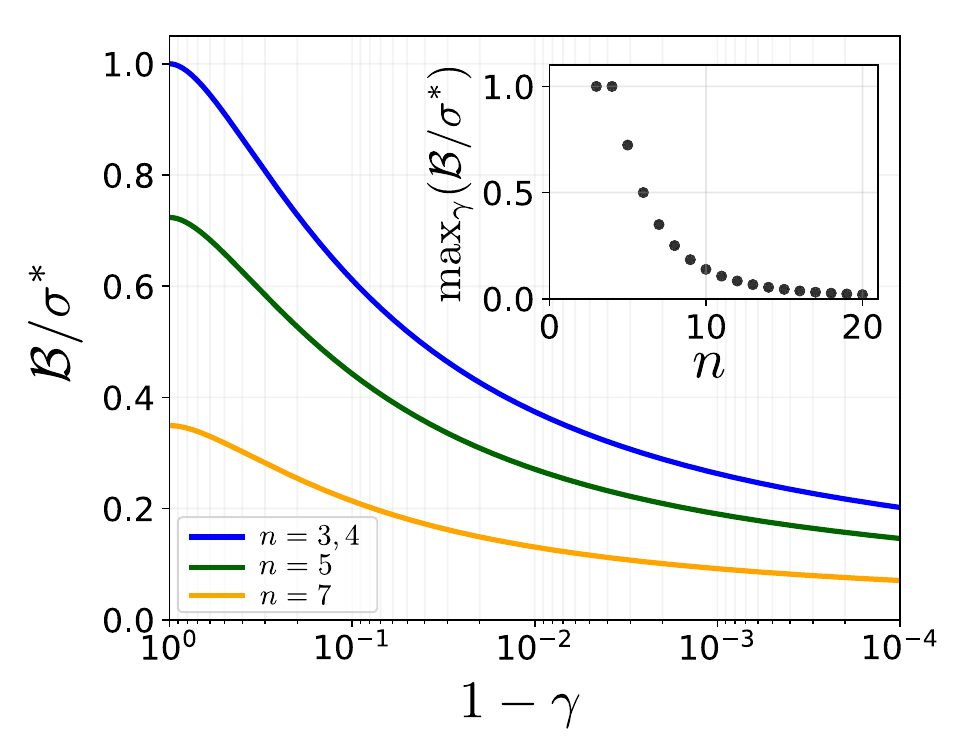}
    \caption{Tightness of the derived bound $\mathcal{B}$ of Eq. \eqref{MJP bound} relative to the entropy production rate $\sigma^*$ for the uniform biased cycle model, see Eq. \eqref{Cycle Tightness}, perturbed along the first Fourier mode. The main panel plots the ratio $\mathcal{B}/\sigma^*$ as a function of the rate asymmetry $1-\gamma$ for cycle networks with $n=3, 4, 5, 7$ states. Note that $n=3$ and $n=4$ give the same curve. (Inset) The global maximum of the tightness ratio, $\max_\gamma (\mathcal{B}/\sigma^*)$, plotted against the number of states $n$.} 
    \label{fig:tightness_cycle}
\end{figure}

Let us consider a perturbation in the form
$\phi_i = \epsilon \cos\left( \delta i \right)$, where $\delta\equiv 2\pi/n$. This is one of the first Fourier mode, corresponding to the slowest relaxation.
Note that $\epsilon \ll 1$ ensures the linear response regime, while normalization can be restored by considering $\mathcal{O}(\epsilon^2)$ terms.
This perturbation evolves as
\begin{equation}
    d_t \phi_i = \epsilon \left[ c_- \sin\left(\delta i\right) - c_+ \cos\left(\delta i\right) \right] ,
\end{equation}
where we defined $c_- \equiv (k_+ - k_-) \sin \delta$ and $c_+ \equiv (k_+ + k_-) (1 - \cos\delta)$, and we used the relation $\cos(\theta\pm \delta) = \cos\theta\cos\delta \mp \sin\theta\sin\delta$.
The squared intrinsic speed is computed here as $(v_{\rm info} (\boldsymbol{p}))^2 =\epsilon^2 (c_-^2 + c_+^2)/2$,
where we used the identities $ \sum_i \sin^2(\delta i)= \sum_i \cos^2(\delta i) =n/2$ and $\sum_i \sin(\delta i)\cos(\delta i) =0$ valid for $n\geq 3$.
Similarly, we obtain $d^2_t D[\boldsymbol{p}||\boldsymbol{p}^*] - 2 (v_{\rm info} (\boldsymbol{p}))^2 = -\epsilon^2 c_-^2 $,
which, as expected, depends on the irreversible rate $c_-$.
We also obtain $\kappa = \frac{1}{2} n (k_+ + k_-)$, $D[\boldsymbol{p}||\boldsymbol{p}^*]=\epsilon^2/4$,  $d_t D[\boldsymbol{p}||\boldsymbol{p}^*]=-\epsilon^2 c_+ /2$, and $\sigma^* =(k_+ -k_-) \ln (k_+ / k_-)$. The equilibrium case corresponds to $k^+=k^-$. The tightness is then, in terms of the asymmetry $\gamma\equiv (k_+ - k_-)/(k_+ + k_-)$, given by
\begin{equation}\label{Cycle Tightness}
\frac{\mathcal{B}}{\sigma^*} = \frac{4 \gamma \sin^2 (2\pi/n)}{n \tanh^{-1}(\gamma)},
\end{equation}
and its behavior is studied numerically in Fig. \ref{fig:tightness_cycle} as a function of $1-\gamma$ and $n$.
We observe that the optimal tightness is obtained in the near-equilibrium limit $\gamma \rightarrow 0$, and it vanishes in the infinite driving limit $\gamma \to 1$.
Both $n=3$ and $n=4$ saturate the inequality $\sigma^*= \mathcal{B}$ with the first Fourier mode when $\gamma \to 0$, while the tightness vanishes $\mathcal{B}/\sigma^* \rightarrow 0$ in the limit of large networks $n\rightarrow \infty$.

\paragraph*{Relaxation gap for Fokker--Planck dynamics.}

The thermodynamic bound \eqref{MJP bound} derived for discrete MJPs can be extended to continuous systems governed by the Fokker-Planck equation $\partial_t P(\boldsymbol{x};t) = -\boldsymbol{\nabla} \cdot (P(\boldsymbol{x};t)\boldsymbol{\nu}(\boldsymbol{x};t))$, where $    \boldsymbol{\nu}(\boldsymbol{x};t) = \boldsymbol{F} (\boldsymbol{x})-T \boldsymbol{\nabla}\ln P(\boldsymbol{x};t)$, and $T$ is the temperature. The steady-state distribution $P^*(\boldsymbol{x})$ is defined by $-\boldsymbol{\nabla} \cdot (P^*(\boldsymbol{x})\boldsymbol{\nu}^*(\boldsymbol{x})) =0$ with $\boldsymbol{\nu}^*(\boldsymbol{x})\equiv \boldsymbol{F} (\boldsymbol{x})-T \boldsymbol{\nabla}\ln P^*(\boldsymbol{x})$.
 The relaxation gap can similarly be used to lower bound the continuous steady-state entropy production rate $\sigma^* = \langle ||\boldsymbol{\nu}^*||^2 \rangle / T =(1/T) \int d\boldsymbol{x}P^*(\boldsymbol{x})\| \boldsymbol{\nu}^*(\boldsymbol{x})\|^2 $. We find that the corresponding thermodynamic bound in the weak-perturbation limit takes the form (see derivation in the SM),
\begin{equation}\label{FP bound}
    \sigma^{*} \ge \frac{\left[d_{t}^{2}D[P||P^{*}] - 2(v_{\rm info} (P))^2\right]^{2}}{4\mu \, (v_{\rm info} (P))^2},
\end{equation}
where $D[P||P^{*}]\equiv \int d\boldsymbol{x} P(\boldsymbol{x};t)\ln [P(\boldsymbol{x};t)/P^*(\boldsymbol{x})]$, $(v_{\rm info} (P))^2 \equiv \int d\boldsymbol{x} P(\boldsymbol{x};t) (d_t \ln P(\boldsymbol{x};t))^2$, and the parameter $\mu$ is defined by the maximum spatial gradient of the perturbation,
\begin{equation}
    \mu \equiv T \max_{\boldsymbol{x}} ||\boldsymbol{\nabla} \Phi(\boldsymbol{x};t)||^2,
\end{equation}
where $\Phi(\boldsymbol{x};t)$ is introduced as $P(\boldsymbol{x};t)= P^*(\boldsymbol{x})\exp[\Phi(\boldsymbol{x};t)]$. As in the derivation of the thermodynamic uncertainty relation~\cite{dechant2018current,otsubo2020estimating, dechant2022geometric, ito2022information}, this quantity can also be expressed in terms of a short-time conditional variance (see also the SM).
The structural difference of this continuous bound \eqref{FP bound} from the discrete one in \eqref{MJP bound} originates from the lack of an intrinsic spatial scale as continuous space permits perturbations of arbitrary sharpness.
We note that this bound, Eq. \eqref{FP bound}, is saturated in a linear model in the near-equilibrium regime, see the SM.

\paragraph*{Derivation of Eq. \eqref{MJP bound}.}

For two positive numbers, the logarithmic mean is less than or equal to the arithmetic mean~\cite{bhatia2008logarithmic}. In the present notation, this gives $J^*_{ij}/F^*_{ij} \leq A^*_{ij}/2$~\cite{shiraishi2018speed,nagayama2025infinite}. This inequality implies a lower bound for the entropy production as
\begin{equation}\label{log inequality}
    \sigma^* \geq \sum_{i,j} \frac{(J^*_{ij})^2}{A^*_{ij}} ,
\end{equation}
and it becomes tight in the near-equilibrium limit where the net currents are much smaller than the dynamical activity, $\vert J^*_{ij} \vert \ll \vert A^*_{ij} \vert$, as can be seen from a first-order expansion of Eq. \eqref{EP}.
We build on this logarithmic inequality and derive a connection with information-theoretic measures as follows. 

Let us first use the anti-symmetric property $J^*_{ij}=-J^*_{ji}$ to rewrite Eq. \eqref{symmetry breaking} as
\begin{equation}
    d^2_t D [\boldsymbol{p} || \boldsymbol{p}^*] - 2 \left( v_{\rm info} (\boldsymbol{p}) \right)^2 = \frac{1}{2} \sum_{i,j}  J^*_{ij} \left( \phi_i d_t\phi_j - \phi_j d_t\phi_i \right).
\end{equation}

Applying the Cauchy-Schwarz inequality, and then the logarithmic inequality of Eq. \eqref{log inequality}, we obtain
\begin{align}
         &\left[ \sum_{i,j}  J^*_{ij}  \left( \phi_i d_t\phi_j - \phi_j d_t\phi_i \right) \right]^2  \nonumber\\
         \leq&  
        \left( \sum_{i,j|i\neq j}  \frac{(J^*_{ij})^2}{A^*_{ij}} \right)  \sum_{i,j|i\neq j} A^*_{ij} \left( \phi_i d_t\phi_j - \phi_j d_t\phi_i \right)^2  \nonumber\\
        \leq&  2 \sigma^* \sum_{i,j} A^*_{ij} \left[ \left( \phi_i d_t\phi_j  \right)^2 - \phi_i\phi_j (d_t\phi_i) (d_t\phi_j)  \right] ,
\end{align}
where we used the $A^*_{ij} > 0$ property for $i\neq j$ to multiply and divide by $(A^*_{ij})^{1/2}$ and $J^*_{ii}=0$.

Finally we use the bound $2 \kappa p_i^* p_j^* \geq A^*_{ij}$ to decouple the cross-terms $\phi_id_t\phi_j$ using the definition of Eq. \eqref{kappa}, and then the definitions in Eqs. \eqref{def D}-\eqref{symmetry breaking} to identify the macroscopic observables as
\begin{align}\label{max majorization}
    &\sum_{i,j} A^*_{ij} \left[ \left( \phi_i d_t\phi_j  \right)^2 - \phi_i\phi_j (d_t\phi_i) (d_t\phi_j)  \right]  \nonumber\\
    \leq& (2\kappa) \sum_{i,j} p^*_i p^*_j \left[ \left( \phi_i d_t\phi_j  \right)^2 - \phi_i\phi_j (d_t\phi_i) (d_t\phi_j)  \right] \nonumber \\
    =&(2\kappa) \left[  2 D[\boldsymbol{p} || \boldsymbol{p}^*] (v_{\rm info} (\boldsymbol{p}))^2 - \left( d_t D[\boldsymbol{p} || \boldsymbol{p}^*] \right)^2 \right],
\end{align}
which completes the derivation.

\paragraph*{Discussion.}
In conclusion, we have established that the normalized squared relaxation gap provides a lower bound on the NESS entropy production rate. This geometric approach may offer a practical way to infer dissipation in the NESS directly from ensemble relaxation dynamics, without requiring an explicit equilibrium reference. Crucially, although our framework assumes weak perturbations around the steady state, this assumption is less restrictive than a near-equilibrium approximation. Indeed, the derived bounds remain valid for NESSs arbitrarily far from equilibrium, provided that the system is considered in the vicinity of the NESS.

The inequality becomes less tight in high-dimensional systems because the use of a single global mixing rate $\kappa$ amounts to a maximization over relaxation channels. It would be interesting to incorporate local network topology, which may avoid the global maximization step and lead to tighter thermodynamic bounds for complex nonconservative systems.

\begin{acknowledgments}
S.I.\ is supported by JSPS KAKENHI Grants No.~22H01141, No.~23H00467, and No.~24H00834, JST ERATO Grant No.~JPMJER2302 and UTEC-UTokyo FSI Research Grant Program.
\end{acknowledgments}
\bibliography{biblio}

@PREAMBLE{
 "\providecommand{\noopsort}[1]{}" 
 # "\providecommand{\singleletter}[1]{#1}%" 
}

@article{tome2025irreversible,
  title={Irreversible thermodynamics and Glansdorff--Prigogine principle derived from stochastic thermodynamics},
  author={Tom{\'e}, T{\^a}nia and Oliveira, M{\'a}rio J de},
  journal={Journal of Statistical Mechanics: Theory and Experiment},
  volume={2025},
  number={6},
  pages={063202},
  year={2025},
  publisher={IOP Publishing}
}

@article{esposito2010three,
  title={Three faces of the second law. I. Master equation formulation},
  author={Esposito, Massimiliano and Van den Broeck, Christian},
  journal={Physical Review E—Statistical, Nonlinear, and Soft Matter Physics},
  volume={82},
  number={1},
  pages={011143},
  year={2010},
  publisher={APS}
}

@book{seifert2025stochastic,
  title={Stochastic thermodynamics},
  author={Seifert, Udo},
  volume={140},
  year={2025},
  publisher={Cambridge University Press Cambridge}
}

@article{schnakenberg1976network,
  title={Network theory of microscopic and macroscopic behavior of master equation systems},
  author={Schnakenberg, J{\"u}rgen},
  journal={Reviews of Modern physics},
  volume={48},
  number={4},
  pages={571},
  year={1976},
  publisher={APS}
}

@article{barato2017coherence,
  title={Coherence of biochemical oscillations is bounded by driving force and network topology},
  author={Barato, Andre C and Seifert, Udo},
  journal={Physical Review E},
  volume={95},
  number={6},
  pages={062409},
  year={2017},
  publisher={APS}
}

@article{ohga2023thermodynamic,
  title={Thermodynamic bound on the asymmetry of cross-correlations},
  author={Ohga, Naruo and Ito, Sosuke and Kolchinsky, Artemy},
  journal={Physical Review Letters},
  volume={131},
  number={7},
  pages={077101},
  year={2023},
  publisher={APS}
}

@article{uhl2019affinity,
  title={Affinity-dependent bound on the spectrum of stochastic matrices},
  author={Uhl, Matthias and Seifert, Udo},
  journal={Journal of Physics A: Mathematical and Theoretical},
  volume={52},
  number={40},
  pages={405002},
  year={2019},
  publisher={IOP Publishing}
}

@article{de1975glansdorff,
  title={The Glansdorff-Prigogine thermodynamic stability criterion in the light of Lyapunov's theory},
  author={de Sobrino, Luis},
  journal={Journal of Theoretical Biology},
  volume={54},
  number={2},
  pages={323--333},
  year={1975},
  publisher={Elsevier}
}

@article{schlogl1971stability,
  title={On stability of steady states},
  author={Schl{\"o}gl, F},
  journal={Zeitschrift f{\"u}r Physik A Hadrons and nuclei},
  volume={243},
  number={4},
  pages={303--310},
  year={1971},
  publisher={Springer}
}

@article{dechant2022geometric,
  title={Geometric decomposition of entropy production into excess, housekeeping, and coupling parts},
  author={Dechant, Andreas and Sasa, Shin-ichi and Ito, Sosuke},
  journal={Physical Review E},
  volume={106},
  number={2},
  pages={024125},
  year={2022},
  publisher={APS}
}

@article{glansdorff1964general,
  title={On a general evolution criterion in macroscopic physics},
  author={Glansdorff, Paul and Prigogine, Ilya},
  journal={Physica},
  volume={30},
  number={2},
  pages={351--374},
  year={1964},
  publisher={Elsevier}
}

@article{dechant2018current,
  title={Current fluctuations and transport efficiency for general Langevin systems},
  author={Dechant, Andreas and Sasa, Shin-ichi},
  journal={Journal of Statistical Mechanics: Theory and Experiment},
  volume={2018},
  number={6},
  pages={063209},
  year={2018},
  publisher={IOP Publishing and SISSA}
}

@article{aslyamov2025nonequilibrium,
  title={Nonequilibrium fluctuation-response relations: From identities to bounds},
  author={Aslyamov, Timur and Ptaszy{\'n}ski, Krzysztof and Esposito, Massimiliano},
  journal={Physical Review Letters},
  volume={134},
  number={15},
  pages={157101},
  year={2025},
  publisher={APS}
}

@article{owen2020universal,
  title={Universal thermodynamic bounds on nonequilibrium response with biochemical applications},
  author={Owen, Jeremy A and Gingrich, Todd R and Horowitz, Jordan M},
  journal={Physical Review X},
  volume={10},
  number={1},
  pages={011066},
  year={2020},
  publisher={APS}
}

@article{sivak2012thermodynamic,
  title={Thermodynamic metrics and optimal paths},
  author={Sivak, David A and Crooks, Gavin E},
  journal={Physical review letters},
  volume={108},
  number={19},
  pages={190602},
  year={2012},
  publisher={APS}
}

@article{maes2020frenesy,
  title={Frenesy: Time-symmetric dynamical activity in nonequilibria},
  author={Maes, Christian},
  journal={Physics Reports},
  volume={850},
  pages={1--33},
  year={2020},
  publisher={Elsevier}
}

@article{nicholson2020time,
  title={Time--information uncertainty relations in thermodynamics},
  author={Nicholson, Schuyler B and Garc{\'\i}a-Pintos, Luis Pedro and del Campo, Adolfo and Green, Jason R},
  journal={Nature Physics},
  volume={16},
  number={12},
  pages={1211--1215},
  year={2020},
  publisher={Nature Publishing Group UK London}
}

@article{xu2025thermodynamic,
  title = {Thermodynamic Geometric Constraint on the Spectrum of Markov Rate Matrices},
  author = {Xu, Guo-Hua and Kolchinsky, Artemy and Delvenne, Jean-Charles and Ito, Sosuke},
  journal = {Phys. Rev. Lett.},
  volume = {135},
  issue = {25},
  pages = {257102},
  numpages = {8},
  year = {2025},
  month = {Dec},
  publisher = {American Physical Society},
 }

@article{otsubo2020estimating,
  title={Estimating entropy production by machine learning of short-time fluctuating currents},
  author={Otsubo, Shun and Ito, Sosuke and Dechant, Andreas and Sagawa, Takahiro},
  journal={Physical Review E},
  volume={101},
  number={6},
  pages={062106},
  year={2020},
  publisher={APS}
}

@article{shiraishi2019information,
  title={Information-theoretical bound of the irreversibility in thermal relaxation processes},
  author={Shiraishi, Naoto and Saito, Keiji},
  journal={Physical review letters},
  volume={123},
  number={11},
  pages={110603},
  year={2019},
  publisher={APS}
}

@article{kolchinsky2026generalized,
  title={Generalized free energy and excess/housekeeping decomposition in nonequilibrium systems: From large deviations to thermodynamic speed limits},
  author={Kolchinsky, Artemy and Dechant, Andreas and Yoshimura, Kohei and Ito, Sosuke},
  journal={Physical Review Research},
  volume={8},
  number={2},
  pages={023025},
  year={2026},
  publisher={APS}
}

@article{crooks2007measuring,
  title={Measuring thermodynamic length},
  author={Crooks, Gavin E},
  journal={Physical Review Letters},
  volume={99},
  number={10},
  pages={100602},
  year={2007},
  publisher={APS}
}

@book{bremaud2013markov,
  title={Markov chains: Gibbs fields, Monte Carlo simulation, and queues},
  author={Br{\'e}maud, Pierre},
  volume={31},
  year={2013},
  publisher={Springer Science \& Business Media}
}

@article{ichiki2013violation,
  title={Violation of detailed balance accelerates relaxation},
  author={Ichiki, Akihisa and Ohzeki, Masayuki},
  journal={Physical Review E—Statistical, Nonlinear, and Soft Matter Physics},
  volume={88},
  number={2},
  pages={020101},
  year={2013},
  publisher={APS}
}

@article{suwa2010markov,
  title={Markov chain Monte Carlo method without detailed balance},
  author={Suwa, Hidemaro and Todo, Synge},
  journal={Physical review letters},
  volume={105},
  number={12},
  pages={120603},
  year={2010},
  publisher={APS}
}

@article{diaconis2000analysis,
  title={Analysis of a nonreversible Markov chain sampler},
  author={Diaconis, Persi and Holmes, Susan and Neal, Radford M},
  journal={Annals of Applied Probability},
  pages={726--752},
  year={2000},
  publisher={JSTOR}
}

@article{turitsyn2011irreversible,
  title={Irreversible Monte Carlo algorithms for efficient sampling},
  author={Turitsyn, Konstantin S and Chertkov, Michael and Vucelja, Marija},
  journal={Physica D: Nonlinear Phenomena},
  volume={240},
  number={4-5},
  pages={410--414},
  year={2011},
  publisher={Elsevier}
}

@article{kaiser2017acceleration,
  title={Acceleration of convergence to equilibrium in Markov chains by breaking detailed balance},
  author={Kaiser, Marcus and Jack, Robert L and Zimmer, Johannes},
  journal={Journal of statistical physics},
  volume={168},
  number={2},
  pages={259--287},
  year={2017},
  publisher={Springer}
}

@article{ito2024geometric,
  title={Geometric thermodynamics for the Fokker--Planck equation: stochastic thermodynamic links between information geometry and optimal transport},
  author={Ito, Sosuke},
  journal={Information geometry},
  volume={7},
  number={Suppl 1},
  pages={441--483},
  year={2024},
  publisher={Springer}
}

@article{prost2009generalized,
  title={Generalized fluctuation-dissipation theorem for steady-state systems},
  author={Prost, Jacques and Joanny, J-F and Parrondo, Juan MR},
  journal={Physical review letters},
  volume={103},
  number={9},
  pages={090601},
  year={2009},
  publisher={APS}
}

@article{mandal2016analysis,
  title={Analysis of slow transitions between nonequilibrium steady states},
  author={Mandal, Dibyendu and Jarzynski, Christopher},
  journal={Journal of Statistical Mechanics: Theory and Experiment},
  volume={2016},
  number={6},
  pages={063204},
  year={2016},
  publisher={IOP Publishing and SISSA}
}

@article{barato2015thermodynamic,
  title={Thermodynamic uncertainty relation for biomolecular processes},
  author={Barato, Andre C and Seifert, Udo},
  journal={Physical review letters},
  volume={114},
  number={15},
  pages={158101},
  year={2015},
  publisher={APS}
}

@article{shiraishi2018speed,
  title = {Speed Limit for Classical Stochastic Processes},
  author = {Shiraishi, Naoto and Funo, Ken and Saito, Keiji},
  journal = {Phys. Rev. Lett.},
  volume = {121},
  issue = {7},
  pages = {070601},
  numpages = {6},
  year = {2018},
  month = {Aug},
  publisher = {American Physical Society},
  }

@article{ito2018stochastic,
  title={Stochastic thermodynamic interpretation of information geometry},
  author={Ito, Sosuke},
  journal={Physical review letters},
  volume={121},
  number={3},
  pages={030605},
  year={2018},
  publisher={APS}
}

@article{ito2020stochastic,
  title={Stochastic time evolution, information geometry, and the Cram{\'e}r-Rao bound},
  author={Ito, Sosuke and Dechant, Andreas},
  journal={Physical Review X},
  volume={10},
  number={2},
  pages={021056},
  year={2020},
  publisher={APS}
}

@article{horowitz2020thermodynamic,
  title={Thermodynamic uncertainty relations constrain non-equilibrium fluctuations},
  author={Horowitz, Jordan M and Gingrich, Todd R},
  journal={Nature Physics},
  volume={16},
  number={1},
  pages={15--20},
  year={2020},
  publisher={Nature Publishing Group UK London}
}

@book{risken1996fokker,
  title={Fokker-planck equation},
  author={Risken, Hannes},
  year={1996},
  publisher={Springer}
}

@article{auconi2025nonequilibrium,
  title={Nonequilibrium relaxation inequality on short timescales},
  author={Auconi, Andrea},
  journal={Physical Review Letters},
  volume={134},
  number={8},
  pages={087104},
  year={2025},
  publisher={APS}
}

@article{kolchinsky2024thermodynamic,
  title={Thermodynamic bound on spectral perturbations, with applications to oscillations and relaxation dynamics},
  author={Kolchinsky, Artemy and Ohga, Naruo and Ito, Sosuke},
  journal={Physical Review Research},
  volume={6},
  number={1},
  pages={013082},
  year={2024},
  publisher={APS}
}

@article{nagayama2025infinite,
  title={Infinite variety of thermodynamic speed limits with general activities},
  author={Nagayama, Ryuna and Yoshimura, Kohei and Ito, Sosuke},
  journal={Physical Review Research},
  volume={7},
  number={1},
  pages={013307},
  year={2025},
  publisher={APS}
}

@book{amari2016information,
  title={Information geometry and its applications},
  author={Amari, Shun-ichi},
  volume={194},
  year={2016},
  publisher={Springer}
}

@article{dechant2021continuous,
  title={Continuous time reversal and equality in the thermodynamic uncertainty relation},
  author={Dechant, Andreas and Sasa, Shin-ichi},
  journal={Physical Review Research},
  volume={3},
  number={4},
  pages={L042012},
  year={2021},
  publisher={APS}
}

@misc{van1983stochastic,
  title={Stochastic processes in physics and chemistry},
  author={Van Kampen, Nicolaas Godfried and Reinhardt, William P},
  year={1983},
  publisher={American Institute of Physics}
}

@article{hatano2001steady,
  title={Steady-state thermodynamics of Langevin systems},
  author={Hatano, Takahiro and Sasa, Shin-ichi},
  journal={Physical review letters},
  volume={86},
  number={16},
  pages={3463},
  year={2001},
  publisher={APS}
}

@article{dechant2020fluctuation,
  title={Fluctuation--response inequality out of equilibrium},
  author={Dechant, Andreas and Sasa, Shin-ichi},
  journal={Proceedings of the National Academy of Sciences},
  volume={117},
  number={12},
  pages={6430--6436},
  year={2020},
  publisher={National Academy of Sciences}
}

@book{chung1997spectral,
  title={Spectral graph theory},
  author={Chung, Fan RK},
  volume={92},
  year={1997},
  publisher={American Mathematical Soc.}
}

@article{ito2022information,
  title={Information geometry, trade-off relations, and generalized Glansdorff--Prigogine criterion for stability},
  author={Ito, Sosuke},
  journal={Journal of Physics A: Mathematical and Theoretical},
  volume={55},
  number={5},
  pages={054001},
  year={2022},
  publisher={IOP Publishing}
}

@article{maes2015revisiting,
  title={Revisiting the Glansdorff--Prigogine criterion for stability within irreversible thermodynamics},
  author={Maes, Christian and Neto{\v{c}}n{\`y}, Karel},
  journal={Journal of Statistical Physics},
  volume={159},
  number={6},
  pages={1286--1299},
  year={2015},
  publisher={Springer}
}

@article{glansdorff1974thermodynamic,
  title={The thermodynamic stability theory of non-equilibrium states},
  author={Glansdorff, Paul and Nicolis, G and Prigogine, I},
  journal={Proceedings of the National Academy of Sciences},
  volume={71},
  number={1},
  pages={197--199},
  year={1974}
}

@article{bhatia2008logarithmic,
  title={The logarithmic mean},
  author={Bhatia, Rajendra},
  journal={Resonance},
  volume={13},
  pages={583--594},
  year={2008},
  publisher={Springer}
}

\end{document}


\preprint{APS/123-QED}

\title{Supplementary Materials for the manuscript "Information-Geometric Signatures of Nonconservative Driving"}


\maketitle

\section{Derivation of Eqs. (7)-(8)}
Consider the perturbation vector definition $p_i = p^*_i e^{\phi_i}$, and use it to rewrite the MJP dynamics Eq. (1) in the main text as
\begin{equation}\label{d_t phi2}
    d_t \phi_i = d_t \ln p_i = \frac{1}{2 p_i} \sum_{j=1}^n (A_{ij} +J_{ij}) 
    = \frac{e^{-\phi_i}}{2 p^*_i} \sum_{j=1}^n (A^*_{ij} +J^*_{ij}) e^{\phi_j} .
\end{equation}
In the small-perturbations limit where $\max_i |\phi_i|\ll 1$ we can expand the exponentials to obtain
\begin{equation}\label{d_t phi2}
    d_t \phi_i = \frac{1}{2 p^*_i} \sum_{j=1}^n (A^*_{ij} +J^*_{ij}) (1+\phi_j -\phi_i) +\mathcal{O}(\phi^2) 
    =  \frac{1}{2 p^*_i} \sum_{j=1}^n (A^*_{ij} +J^*_{ij}) \phi_j  +\mathcal{O}(\phi^2),
\end{equation}
where we used the symmetries $\sum_{j=1}^n A^*_{ij} = \sum_{j=1}^n J^*_{ij} = 0$.

The exponential form of the perturbation means that the normalization is of the form $\langle e^\phi\rangle=1$, and in the small perturbation limit this gives
\begin{equation}
    \langle \phi \rangle = -\frac{1}{2} \langle \phi^2 \rangle  +\mathcal{O}(\langle \phi^3 \rangle) .
\end{equation}
The divergence is then computed as
\begin{equation}
   D[p||p^*] = \langle \phi e^\phi \rangle
   = \langle \phi \rangle + \langle \phi^2 \rangle +\mathcal{O}(\langle \phi^3 \rangle) 
   = \frac{1}{2} \langle \phi^2 \rangle +\mathcal{O}(\langle \phi^3 \rangle) .
\end{equation}
Note that in the last step we could have taken equivalently $D[p||p^*] = -\langle \phi \rangle +\mathcal{O}(\langle \phi^3 \rangle)$, but then we would have had to keep $\mathcal{O}(\phi^2)$ terms in the dynamics $d_t\phi_i$ for the derivations below.

\section{Derivation of the relation between the intrinsic speed and the Kullback-Leibler divergence} \label{klintrinsic}
Consider the  Kullback-Leibler divergence 
\begin{equation}
    D[\boldsymbol{p}(t+\tau)||\boldsymbol{p}(t)] = \sum_i p_i(t+\tau) \ln\left(\frac{p_i(t+\tau)}{p_i(t)}\right) = \langle e^{\phi(t+\tau)} [\phi(t+\tau)-\phi(t)] \rangle .
\end{equation} 
Its time derivative is
\begin{equation}
    d_{\tau} D[\boldsymbol{p}(t+\tau)||\boldsymbol{p}(t)] = \langle (d_{\tau} e^{\phi(t+\tau)}) [\phi(t+\tau)-\phi(t)] \rangle ,
\end{equation} 
where we used the normalization $\langle e^\phi \rangle=1$ in $\langle e^{\phi(t+\tau)}d_{\tau} \phi(t+\tau) \rangle =\langle d_{\tau} e^{\phi(t+\tau)} \rangle = d_{\tau} \langle e^{\phi(t+\tau)} \rangle = 0$. Note that this divergence vanishes to first time order, $d_{\tau}D[\boldsymbol{p}(t+\tau)||\boldsymbol{p}(t)]|_{\tau=0} = 0$.
The second time derivative is
\begin{equation}
    d^2_{\tau} D[\boldsymbol{p}(t+\tau)||\boldsymbol{p}(t)] = \langle (d^2_{\tau} e^{\phi(t+\tau)}) [\phi(t+\tau)-\phi(t)] \rangle +\langle (d_{\tau} e^{\phi(t+\tau)}) (d_{\tau} \phi(t+\tau))  \rangle ,
\end{equation} 
and evaluating it at $\tau=0$ gives
\begin{equation}
    d^2_{\tau} D[\boldsymbol{p}(t+\tau)||\boldsymbol{p}(t)] |_{\tau=0} 
    = \langle (d_t  \phi)^2 e^{\phi} \rangle =  \langle (d_t  \phi)^2 \rangle + \mathcal{O}(\langle \phi (d_t  \phi)^2 \rangle),
\end{equation} 
where the last equality holds to leading order in the small perturbation limit.
We also obtain the intrinsic speed $v_{\rm info}( \boldsymbol{p})$ as
\begin{equation}
    (v_{\rm info} (\boldsymbol{p}(t)) )^2 \equiv 
    \sum_i p_i (d_t \ln p_i)^2 = \langle (d_t  \phi)^2 e^{\phi} \rangle = d^2_{\tau} D[\boldsymbol{p}(t+\tau)||\boldsymbol{p}(t)] |_{\tau=0}.
\end{equation}

\section{The corresponding equilibrium dynamics}

The corresponding equilibrium associated with the NESS is defined as that system having the same steady-state density $\boldsymbol{p}^*$ but satisfying detailed balance. This identification is not unique, and in the continuous Fokker-Planck equation description it is resolved by fixing the temperature \cite{hatano2001steady, dechant2020fluctuation, dechant2021continuous}. Here in the discrete case, following Ref. \cite{bremaud2013markov,kolchinsky2024thermodynamic} we define the corresponding equilibrium rate as
\begin{equation}
    \widetilde{M}_{ij} \equiv \frac{1}{2}\left( M_{ij} + M_{ji} \frac{p^*_i}{p^*_j} \right) = \frac{A^*_{ij}}{2 p^*_j},
\end{equation}
motivated by its conservation of the escape rates $\widetilde{M}_{ii}=M_{ii}$ and steady-state dynamical activity $\widetilde{A_{ij}^*} \equiv \widetilde{M}_{ij}p^*_j +\widetilde{M}_{ji}p^*_i =A^*_{ij}$. Here, $\widetilde{J_{ij}^*} \equiv \widetilde{M}_{ij}p^*_j -\widetilde{M}_{ji}p^*_i=0$.

Subtracting this corresponding equilibrium dynamics from the short-timescale relaxation Eq. (12) of the main text, we get
\begin{align}
    \xi \equiv d^2_t D [p || p^*] - 
    \left. {d^2_t} D [p || p^*] \right|_{\widetilde{M}}
    &= \sum_{i,j} A^*_{ij} \phi_i (d_t \phi_j - \left. d_t \phi_j\right|_{\widetilde{M}})\\
    &=\sum_{i,j,k: j \neq k}  \phi_i \frac{A^*_{ij}J^*_{jk}}{2 p^*_j} \phi_k ,
\end{align}
where $\left. {d^2_t} D [p || p^*] \right|_{\widetilde{M}}$ is the quantity measured by the dynamics using $\widetilde{M}$ and its dynamics for $\boldsymbol{\phi}$ is given by $\left. d_t \phi_j \right|_{\widetilde{M}} =\sum_k (\widetilde{A}^*_{jk}+\widetilde{J^*_{jk}}) \phi_k/(2p^*_j)= \sum_k A^*_{jk} \phi_k/(2p^*_j)$. Here, we used $J^*_{jj}=0$.

From the Cauchy-Schwarz inequality and the logarithmic inequality (24) of the main text, i.e., $\sigma^* \geq \sum_{j,k: j\neq k} (J^*_{jk})^2/A^*_{jk}$, we get the thermodynamic bound
\begin{equation}
   \left[ \sum_{j, k : j\neq k} A^*_{jk}  \left( \frac{ \phi_k}{p_j^*}\sum_i A^*_{ij} \phi_i \right)^2 \right] \sigma^* \geq 4 \xi^2 ,
\end{equation}
which extends the bound of Ref. \cite{auconi2025nonequilibrium} to the MJPs setting.

\section{Numerical analysis of the tightness dependence on the network dimension}

\begin{figure}[htbp]
    \centering
    \begin{subfigure}[b]{0.48\textwidth}
        \centering
        \includegraphics[width=\textwidth]{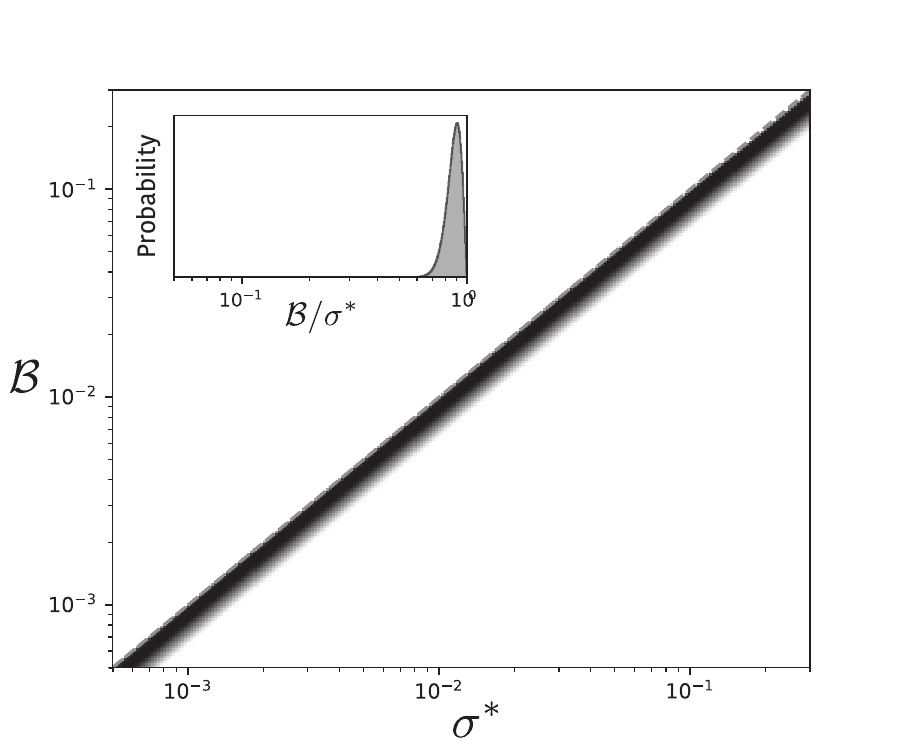}
        \caption{$n=3$}
        \label{fig:tightness_n3}
    \end{subfigure}
    \hfill
    \begin{subfigure}[b]{0.48\textwidth}
        \centering
        \includegraphics[width=\textwidth]{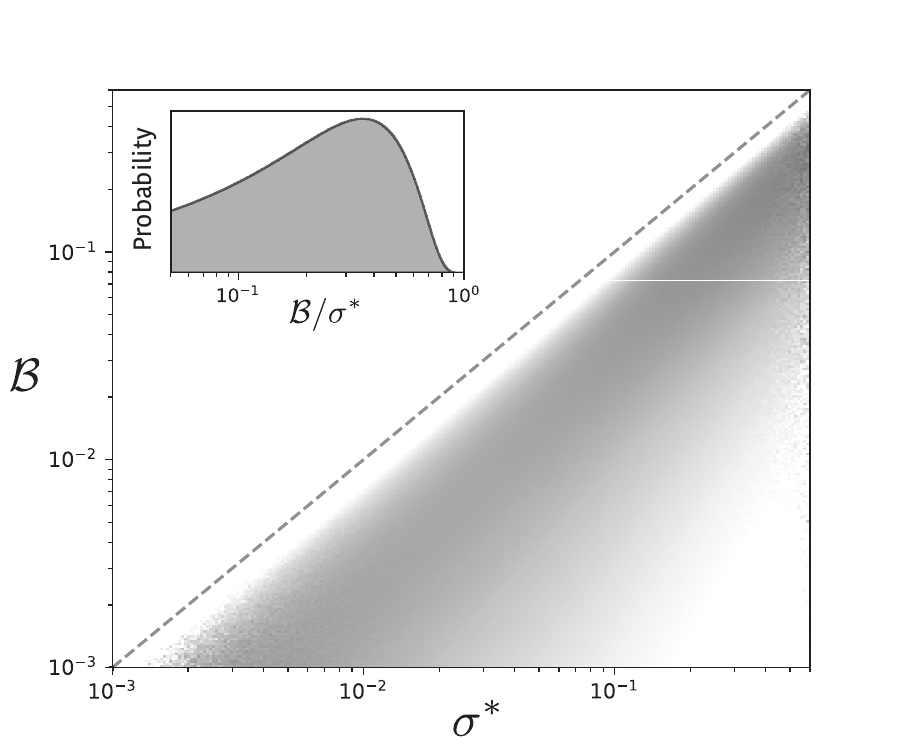}
        \caption{$n=5$}
        \label{fig:tightness_n5}
    \end{subfigure}
    
    \vspace{0.5cm} 
    
    \begin{subfigure}[b]{0.48\textwidth}
        \centering
        \includegraphics[width=\textwidth]{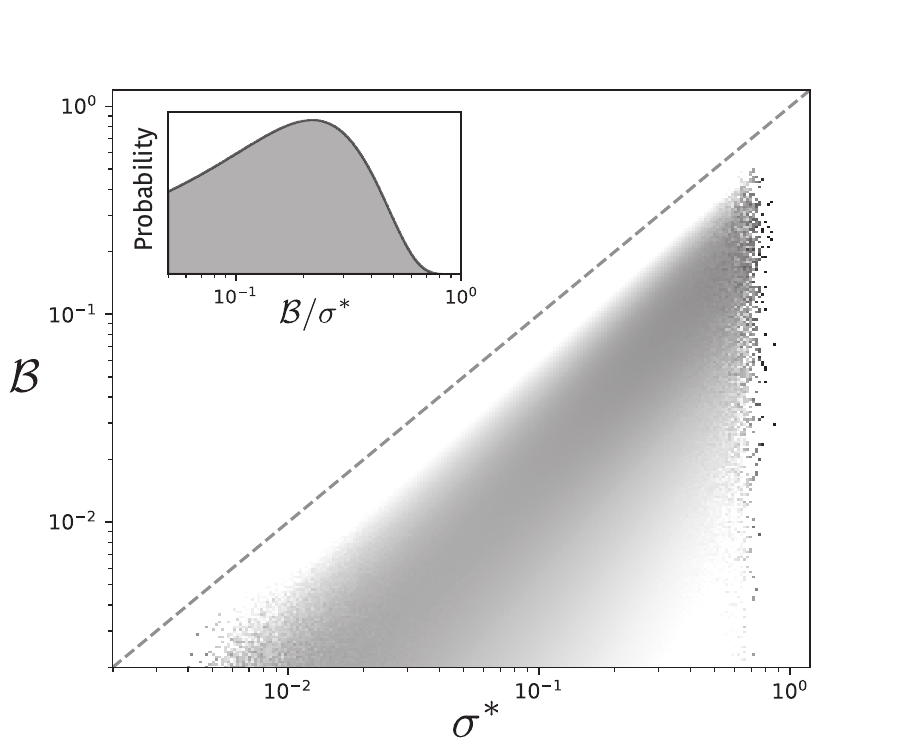}
        \caption{$n=6$}
        \label{fig:tightness_n6}
    \end{subfigure}
    \hfill
    \begin{subfigure}[b]{0.48\textwidth}
        \centering
        \includegraphics[width=\textwidth]{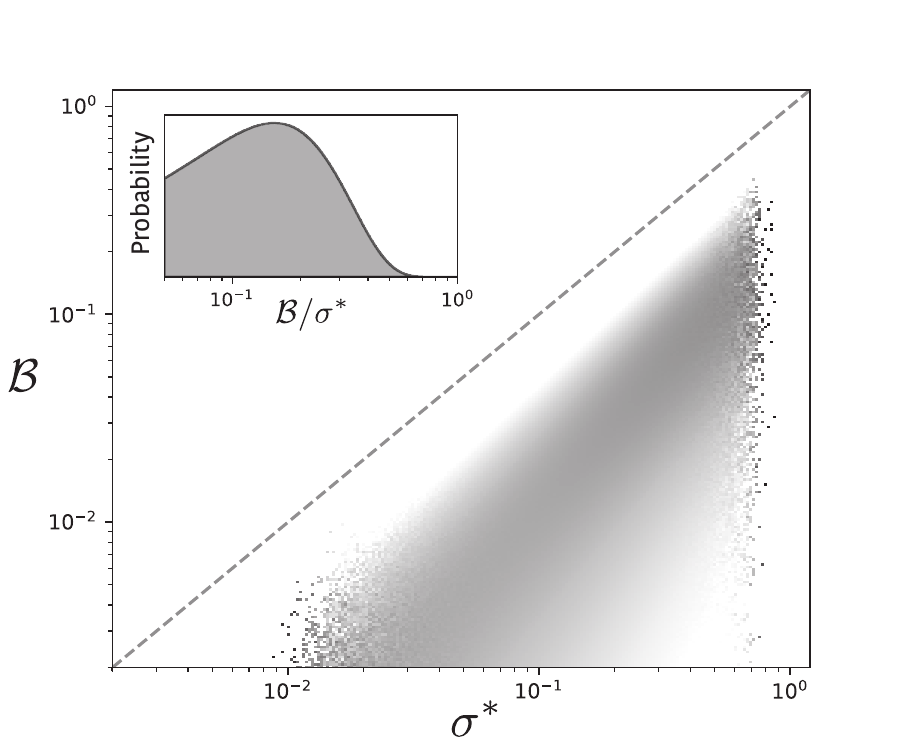}
        \caption{$n=7$}
        \label{fig:tightness_n7}
    \end{subfigure}
    
    \caption{Random systems analysis for various network dimensions. The $n=3$ results are shown to be always relatively tight, as the fully connected triangular topology allows the global maximum mixing rate to accurately proxy local dynamics. For larger dimensions, tightness systematically decreases. This looseness stems from the majorization by maxima step in the bound's derivation, which upper-bounds all local edge activity using the system's single fastest relaxation channel. As the number of edges increases with $n$, this worst-case assumption heavily overestimates the dynamics of slow edges. Finally, we acknowledge that numerical precision limits the evaluation of larger networks, as visible near the plot boundaries of the $n=6$ and $n=7$ cases.}
    \label{fig:supp_random_systems}
\end{figure}

To systematically evaluate the tightness of the thermodynamic bound, we numerically generated and analyzed a large ensemble of random Markov jump processes. As illustrated in Fig.~\ref{fig:supp_random_systems}, this analysis was performed for various network dimensions ($n=3, 5, 6, 7$).

The numerical pipeline is implemented in Python and proceeds by generating random transition matrices for a given number of states $n$. The off-diagonal transition rates of the Markov matrix $M_{ij}$ are sampled independently from a log-normal distribution, while the diagonal elements are set to enforce probability conservation ($\sum_i M_{ij} = 0$). The steady-state distribution $\boldsymbol{p}^*$ is then computed by linear algebra. From this, the steady-state entropy production rate $\sigma^*$ and the maximum mixing rate $\kappa$ are calculated. Systems that fall below numerical precision thresholds are discarded to prevent floating-point instability. We evaluate the bound $\mathcal{B}$ close to the NESS by considering the information-geometric observables in the linear response limit derived in the main text.

The code dynamically updates pre-allocated 2D conditional probability histograms $p(\mathcal{B}|\sigma^*)$ and 1D tightness ratio histograms $p(\mathcal{B}/\sigma^*)$. These are then used to generate the plots shown in Fig.~\ref{fig:supp_random_systems}.

\section{Derivation of Eq. (22)}

The Fokker-Planck equation \cite{risken1996fokker,ito2024geometric} describes the temporal evolution of the probability density $P(\boldsymbol{x};t)$ of a Brownian particle subject to a force field $\boldsymbol{F}(\boldsymbol{x})$. This is a continuity equation for the probability current $\boldsymbol{\nu}(\boldsymbol{x};t)P(\boldsymbol{x};t)$,
\begin{equation}\label{FP1}
    \partial_t P(\boldsymbol{x};t)= -\boldsymbol{\nabla}\cdot \left( \boldsymbol{\nu} (\boldsymbol{x};t) P(\boldsymbol{x};t)\right).
\end{equation}
written in terms of the local mean velocity $\boldsymbol{\nu}(\boldsymbol{x};t)$, which is a sum of terms representing the deterministic drift driven by $\boldsymbol{F}(\boldsymbol{x})$, and the thermal diffusion at temperature $T$,
\begin{equation}\label{FP2}
    \boldsymbol{\nu} (\boldsymbol{x};t)= \boldsymbol{F}(\boldsymbol{x}) -T \boldsymbol{\nabla}\ln P(\boldsymbol{x};t).
\end{equation}
Our analysis relies on the existence of a unique steady-state distribution, $P^*(\boldsymbol{x})$, characterized by $\partial_t P(\boldsymbol{x};t)\vert_{P(\boldsymbol{x};t)= P^*(\boldsymbol{x})}= 0$. Quantities at steady state are marked with an asterisk (*). The system is defined to be in the NESS~\cite{horowitz2020thermodynamic,ito2024geometric} if this invariant density condition is met, yet a non-zero probability current persists, $\boldsymbol{\nu}^* (\boldsymbol{x}) = \boldsymbol{F}(\boldsymbol{x}) -T \boldsymbol{\nabla}\ln P^*(\boldsymbol{x}) \neq \boldsymbol{0}$. 

Let us define the steady-state irreversible entropy production rate as
\begin{equation}\label{Def entropy production}
    \sigma^* = \frac{1}{T} \left\langle || \boldsymbol{\nu}^*||^2 \right\rangle ,
\end{equation}
where $|| \boldsymbol{\nu}^*(\boldsymbol{x};t) ||^2 \equiv \sum_i [\nu_i^*(\boldsymbol{x};t) ]^2$ denotes the Euclidean norm.

\paragraph*{Perturbation.}

Let us introduce a perturbation at time $t=0$ in the form
\begin{equation}
    P(\boldsymbol{x};t)= P^*(\boldsymbol{x})  e^{\Phi(\boldsymbol{x};t) },
\end{equation}
with $\Phi(\boldsymbol{x};t)$ being a smooth perturbation field.
Its time evolution is computed from Eqs. \eqref{FP1}-\eqref{FP2} as
\begin{equation}
    \partial_t \Phi(\boldsymbol{x};t) = T\alpha (\boldsymbol{x};t)- \boldsymbol{\nabla}\Phi(\boldsymbol{x};t) \cdot \boldsymbol{\nu}^*(\boldsymbol{x}) ,
\end{equation}
where we defined
\begin{equation}
    \alpha (\boldsymbol{x};t) \equiv  \nabla^2\Phi (\boldsymbol{x};t) +\boldsymbol{\nabla}\Phi (\boldsymbol{x};t) \cdot \boldsymbol{\nabla}\ln P^*(\boldsymbol{x}) + ||\boldsymbol{\nabla}\Phi(\boldsymbol{x};t)||^2 =\frac{\nabla \cdot (P(\boldsymbol{x};t) \nabla \Phi (\boldsymbol{x};t))}{P(\boldsymbol{x};t)} .
\end{equation}

\paragraph*{The relaxation observable.}

Following \cite{auconi2025nonequilibrium}, we consider the Kullback-Leibler divergence \cite{amari2016information} of the instantaneous density $P(\boldsymbol{x};t) $ from the steady state $P^*(\boldsymbol{x}) $,
\begin{equation}
    D[P(t)||P^*] \equiv \int d\boldsymbol{x} \, P(\boldsymbol{x};t) \ln \left( \frac{P(\boldsymbol{x};t)}{P^*(\boldsymbol{x})} \right) = \left\langle e^\Phi  \Phi \right\rangle,
\end{equation}
where the brackets denote expectations with respect to the steady-state density $\langle f\rangle = \int d\boldsymbol{x} \, P^* (\boldsymbol{x})f(\boldsymbol{x})$ for any test function $f(\boldsymbol{x})$.
Its time derivative is computed as 
\begin{equation}\label{stability FP}
    d_t D[P(t)||P^*] = -T \left\langle e^\Phi ||\boldsymbol{\nabla} \Phi ||^2  \right\rangle \leq 0,
\end{equation}
where integration by parts was performed assuming that $P(\boldsymbol{x};t)$ vanishes fast enough at infinity, and the steady state property $\boldsymbol{\nabla}\cdot (p^*\boldsymbol{\nu}^*) =0$ was used.
Eq. \eqref{stability FP} proves that the steady state is stable also for large perturbations beyond the weak limit considered in Ref. \cite{auconi2025nonequilibrium}.

The second time derivative is computed as
\begin{equation}
    d^2_t D[P||P^*] = T \left\langle e^\Phi (\partial_t \Phi) \left( 2\alpha -||\boldsymbol{\nabla}\Phi||^2 \right) \right\rangle ,
\end{equation}
where $P(\boldsymbol{x};t)$ was assumed to vanish fast enough at infinity.
In the weak perturbation limit, the term $||\boldsymbol{\nabla}\Phi(\boldsymbol{x};t)||^2$ is negligible compared with $\alpha(\boldsymbol{x};t)$, and the result of Ref. \cite{auconi2025nonequilibrium} is recovered.

\paragraph*{Intrinsic speed.}
The intrinsic speed is given by
\begin{equation}\label{line element FP}
    (v_{\text{info}}(P))^2 \equiv \int d\boldsymbol{x}  P(\boldsymbol{x};t) (d_t \ln P(\boldsymbol{x};t))^2 =  \left\langle e^{\Phi}  (\partial_t \Phi)^2 \right\rangle =  \left\langle (\partial_t \Phi)^2 \right\rangle + \mathcal{O}(\langle \Phi (d_t  \Phi)^2 \rangle).
\end{equation}
As in the discrete case (Sec.~\ref{klintrinsic}), this quantity can be expressed using the second derivative of the Kullback–Leibler divergence
\begin{equation}
    D[P(t+\tau)||P(t)] \equiv \int d\boldsymbol{x} \, P(\boldsymbol{x}; t+\tau) \ln \left( \frac{P(\boldsymbol{x}; t+\tau)}{P(\boldsymbol{x}; t)} \right) ,
\end{equation}
as follows:
\begin{equation}
    d_{\tau}^2 D[P(t+\tau)||P(t)] \vert_{\tau =0} = \left\langle e^{\Phi}  (\partial_t \Phi)^2 \right\rangle  = (v_{\text{info}}(P))^2.
\end{equation}

\paragraph*{Weak perturbation limit.}
In the weak perturbation limit where uniformly $\Phi (\boldsymbol{x};t) \rightarrow 0$, we consider this relation valid to leading order,
\begin{equation}
    \frac{1}{2}d^2_t D[P||P^*] -(v_{\text{info}}(P))^2 = \left\langle (\partial_t \Phi)  \boldsymbol{\nabla}\Phi  \cdot \boldsymbol{\nu}^*  \right\rangle .
\end{equation}
Although this expression corresponds to $(1/2) d^2_t D [\boldsymbol{p} || \boldsymbol{p}^*] - \left( v_{\rm info} (\boldsymbol{p}) \right)^2 = (1/4) \sum_{i,j}  J^*_{ij} \left( \phi_i d_t\phi_j - \phi_j d_t\phi_i \right)$ for the discrete state, we cannot use $\kappa$ to evaluate the inequality unlike in the discrete-state case. Therefore, we consider a different inequality.
From the absolute value, maximum value, and Cauchy-Schwarz majorizations, we find
\begin{equation}
    \left\langle (\partial_t \Phi) \boldsymbol{\nabla}\Phi \cdot \boldsymbol{\nu}^* \right\rangle^2 \leq \left\langle (\partial_t \Phi)^2 ||\boldsymbol{\nabla}\Phi||^2\right\rangle \left\langle ||\boldsymbol{\nu}^* ||^2 \right\rangle \\
\leq \left\langle (\partial_t \Phi)^2 \right\rangle \left( \max_{\boldsymbol{x}} ||\boldsymbol{\nabla}\Phi (\boldsymbol{x};t)||^2  \right) \left\langle ||\boldsymbol{\nu}^* ||^2 \right\rangle.
\end{equation}
From the definitions of Eqs. \eqref{Def entropy production}-\eqref{line element FP} we can rewrite this as a thermodynamic bound on the steady-state entropy production rate
\begin{equation}\label{FP_bound_appendix}
     \sigma^* \geq \frac{\left( d^2_t D[P||P^*] - 2 (v_{\text{info}}(P))^2 \right)^2 }{4 \mu \, (v_{\text{info}}(P))^2} \equiv \mathcal{B}',
\end{equation}
where
\begin{equation}\label{mu appendix}
    \mu \equiv  T  \max_{\boldsymbol{x}} ||\boldsymbol{\nabla}\Phi(\boldsymbol{x};t)||^2,
\end{equation}
valid in the weak perturbation limit. As in the derivation of thermodynamic uncertainty relations~\cite{otsubo2020estimating,dechant2022geometric,ito2022information}, this term $\mu$ can also be obtained using a short-time conditional variance
\begin{equation}\label{mu appendix2}
    \mu = \max_{\boldsymbol{x}} \lim_{\Delta t \to 0}\frac{{\rm Var}[\Delta \Phi] \vert_{\boldsymbol{x}}}{2 \Delta t},
\end{equation}
where ${\rm Var}[\Delta \Phi] \vert_{\boldsymbol{x}}$ is the variance of $\Delta \Phi\equiv \Phi(\boldsymbol{x}(t+\Delta t);t+\Delta t) -\Phi(\boldsymbol{x}(t);t)$ with fixed $\boldsymbol{x}(t)=\boldsymbol{x}$. We can verify this as follows. Using the Langevin description $\Delta \boldsymbol{x}=\boldsymbol{F}(\boldsymbol{x}(t)) \Delta t +\sqrt{2T}\Delta \boldsymbol{W}$ and applying the chain rule $\Delta \Phi = (\partial_t\Phi(\boldsymbol{x}(t);t))\Delta t +\boldsymbol{\nabla}\Phi(\boldsymbol{x}(t);t) \cdot \Delta \boldsymbol{x}$, the variance of $\Delta \Phi$ for the Brownian motion increment $\Delta \boldsymbol{W}$ with fixed $\boldsymbol{x}(t)=\boldsymbol{x}$ is calculated as ${\rm Var}[\Delta \Phi] \vert_{\boldsymbol{x}} = 2T \|\boldsymbol{\nabla}\Phi(\boldsymbol{x};t) \|^2 \Delta t + \mathcal{O}((\Delta t)^{3/2})$. Here, we used $\mathbb{E}[(\Delta W_i )(\Delta W_j)] = \delta_{ij} \Delta t$ where $\mathbb{E}[\cdot]$ stands for the expected value.
Taking the limit $\Delta t \to 0$, we obtain Eq.~(\ref{mu appendix2}).


\subsection*{Saturation in a basic linear model}

We now apply the continuous bound \eqref{FP_bound_appendix} in a basic linear model with a periodic spatial perturbation field \cite{auconi2025nonequilibrium}, and show that this saturates the bound.
The periodic spatial perturbation field is
\begin{equation}
\Phi(\boldsymbol{x};t) = \epsilon \sin(\boldsymbol{k} \cdot \boldsymbol{x} + \varphi) + \eta,
\end{equation}
with $\epsilon \ll 1$, $\boldsymbol{k} = k \boldsymbol{n}_{\theta}$, and $\boldsymbol{n}_{\theta} \equiv [\cos(\theta), \sin(\theta)]$. The constant $\eta$ is chosen to enforce normalization, $\int d\boldsymbol{x}P^*(\boldsymbol{x})e^{\Phi(\boldsymbol{x};t)}=1$. Substituting this perturbation into Eq. \eqref{mu appendix} gives $\mu = T \epsilon^2 k^2$. 

Applying the slowly varying limit ($k \to 0$) to the kinematic observables $d_t^2 D[P||P^*]$ and $v_{\text{info}}(P)$, and substituting the resulting expansions into the bound $\mathcal{B}'$, the amplitude and wavenumber dependencies $\epsilon^4 k^4$ cancel out. Selecting the optimal phase $\varphi = 0$ leaves a purely directional bound,
\begin{equation}
\sigma^* \ge \mathcal{B}' = \frac{\left\langle (\boldsymbol{n}_{\theta} \cdot \boldsymbol{\nu^*})\left[\boldsymbol{n}_{\theta}\cdot \left(T\boldsymbol{\nabla} \ln P^* -\boldsymbol{\nu^*} \right)\right] \right\rangle^2}{T \langle \left[\boldsymbol{n}_{\theta}\cdot \left(T\boldsymbol{\nabla} \ln P^* -\boldsymbol{\nu^*} \right)\right]^2 \rangle} .
\end{equation}

We consider a two-dimensional hierarchical system described by the coordinates $\boldsymbol{x}\equiv(x,y)$, and whose drift is $\boldsymbol{F}(\boldsymbol{x}) = [-\beta x , \beta'(x-y) ]$, with $\beta > 0$ and $\beta' > 0$.
By defining the dimensionless asymmetry parameter $\delta \equiv \beta'/(\beta + \beta')$, and the combination $s =(1-\delta+\delta^2)/\delta$, let us rewrite the steady-state expressions derived in \cite{auconi2025nonequilibrium} as
\begin{align}
\boldsymbol{\nabla} \ln P^* (\boldsymbol{x};t)&= C [sx-\delta y, -\delta x+y] , \\
\boldsymbol{\nu}^* (\boldsymbol{x};t)&= \frac{\beta\delta}{s-\delta^2} [\delta x - y, sx - \delta y] ,
\end{align}
where the normalization prefactor is $C \equiv -\beta/[T(s-\delta^2)]$. 
Also note that the steady-state entropy production for this model is $\sigma^* =\beta \delta^2/(1-\delta)$.

Let us now consider the limit of weak driving, $\delta \rightarrow 0$. In this limit, $\boldsymbol{n}_{\theta} \cdot T \boldsymbol{\nabla} \ln P^* = -\beta \cos(\theta) x + \mathcal{O}(\delta)$, $\boldsymbol{n}_{\theta} \cdot \boldsymbol{\nu^*} = \delta \beta \sin(\theta) x + \mathcal{O}(\delta^2)$, and $\sigma^* = \beta \delta^2 + \mathcal{O}(\delta^3)$ hold. Using the spatial variance $\langle x^2 \rangle = T/\beta$, the scaling of the lower bound in the weak driving limit reads
\begin{equation}
    \mathcal{B}' = \beta \delta^2 \sin^2(\theta) + \mathcal{O}(\delta^3),
\end{equation}
so that maximizing the lower bound over all valid directions $\theta$ yields its saturation,
\begin{equation}
    \frac{\sigma^*}{\max_{\theta}\mathcal{B}'} = 1 + \mathcal{O}(\delta).
\end{equation}

\bibliography{biblio}